\documentclass[twocolumn,aps,english,prl,nofootinbib,preprintnumbers]{revtex4}
\usepackage{mathrsfs}
\usepackage{hyperref}
\usepackage{graphicx}
\usepackage{amsmath, amssymb}
\usepackage{babel}
\usepackage{color}
\usepackage{slashed}

\def\beqn{\begin{eqnarray}}
\def\eeqn{\end{eqnarray}}
\def\barr{\begin{array}}
\def\earr{\end{array}}
\def\btab{\begin{tabular}}
\def\etab{\end{tabular}}
\def\bite{\begin{itemize}}
\def\eite{\end{itemize}}
\def\bcen{\begin{center}}
\def\ecen{\end{center}}

\begin{document}

\title{
High-precision determination of the {\boldmath$K_{e3}$} radiative corrections
}

\author{Chien-Yeah Seng$^{1}$}
\author{Daniel Galviz$^{1}$}
\author{Mikhail Gorchtein$^{2,3,4}$}
\author{Ulf-G. Mei{\ss}ner$^{1,5,6}$}

\affiliation{$^{1}$Helmholtz-Institut f\"{u}r Strahlen- und Kernphysik and Bethe Center for Theoretical Physics,\\
	Universit\"{a}t Bonn, 53115 Bonn, Germany}
\affiliation{$^{2}$Helmholtz Institute Mainz, D-55099 Mainz, Germany}
\affiliation{$^{3}$GSI Helmholtzzentrum f\"ur Schwerionenforschung, 64291 Darmstadt, Germany}
\affiliation{$^{4}$Johannes Gutenberg University, D-55099 Mainz, Germany}
\affiliation{$^{5}$Institute for Advanced Simulation, Institut f\"ur Kernphysik and J\"ulich Center for Hadron Physics,
 Forschungszentrum J\"ulich, 52425 J\"ulich, Germany}
\affiliation{$^{6}$Tbilisi State  University,  0186 Tbilisi, Georgia}

\date{\today}

\begin{abstract}
We report a high-precision calculation of the Standard Model electroweak radiative corrections in the $K\to \pi
e^+\nu(\gamma)$ decay as a part of the combined theory effort to understand the existing anomaly in the determinations
of $V_{us}$. Our new analysis features a chiral resummation of the large infrared-singular terms in the radiative
corrections and a well-under-control strong interaction uncertainty based on the most recent lattice QCD inputs.
While being consistent with the current state-of-the-art results obtained from chiral perturbation theory, we reduce
the existing theory uncertainty from $10^{-3}$ to $10^{-4}$. Our result suggests that the Standard Model
electroweak effects cannot account for the $V_{us}$ anomaly. 
\end{abstract}

\maketitle


An interesting anomaly has recently been observed in $V_{us}$, which is a top-row element of the
Cabibbo-Kobayashi-Maskawa (CKM) matrix~\cite{Cabibbo:1963yz,Kobayashi:1973fv} in the Standard Model (SM) of
particle physics. The measured values of this matrix element stem from two different channels of kaon decay,
$K\to \mu\nu(\gamma)$ ($K_{\mu 2}$) and $K\to\pi l^+\nu(\gamma)$ ($K_{l3}$), and show a disagreement at the
$\sim 3\sigma$ level~\cite{Zyla:2020zbs}:
\begin{eqnarray}
|V_{us}|&=&0.2252(5)\:\:(K_{\mu 2})~,\nonumber\\
&=&0.2231(7)\:\:(K_{l3})~,
\end{eqnarray}
which may hint to the existence of physics beyond the Standard Model (BSM). The value obtained from the $K_{l3}$ decay is particularly interesting because it also leads to a violation of the
top-row CKM unitarity at $(3-5)\sigma$ upon combining with the most recent updates of
$V_{ud}$~\cite{Seng:2018yzq,Czarnecki:2019mwq,Seng:2020wjq,Shiells:2020fqp}, depending on the amount of nuclear
uncertainties assigned to the latter~\cite{Seng:2018qru,Gorchtein:2018fxl}. However, despite of an active discussion about the possible BSM origin of the $K_{\mu 2}$--$K_{l3}$
discrepancy~\cite{Belfatto:2019swo,Tan:2019yqp,Grossman:2019bzp,Coutinho:2019aiy,Cheung:2020vqm,Crivellin:2020lzu,Endo:2020tkb,Capdevila:2020rrl,Kirk:2020wdk}, the current significance level is not yet sufficient to claim a discovery. One of the main obstacles is the large hadronic uncertainty in
the electroweak radiative corrections (EWRC), which are the focus of this work.

Among the many studies of the EWRC in $K_{l3}$~\cite{Ginsberg:1966zz,Ginsberg:1968pz,Ginsberg:1969jh,Ginsberg:1970vy,Becherrawy:1970ah,Bytev:2002nx,Andre:2004tk,Garcia:1981it,JuarezLeon:2010tj,Torres:2012ge,Neri:2015eba}, the standard inputs in
global analyses~\cite{Antonelli:2010yf,Cirigliano:2011ny} are based on chiral perturbation theory
(ChPT) which is the low-energy effective field theory of Quantum Chromodynamics (QCD). Within this framework, the
``short-distance'' electroweak corrections are isolated as a constant factor, while the ``long-distance''
electromagnetic corrections are calculated up to $\mathcal{O}(e^2p^2)$~\cite{Cirigliano:2001mk,Cirigliano:2004pv,Cirigliano:2008wn}, with $e$ the electric charge and  $p$ a small momentum/meson mass.
The estimated theory uncertainties in these calculations are of the order $10^{-3}$, and originate
from: (1) the neglected contributions at $\mathcal{O}(e^2p^4)$, and (2) the contributions from 
non-perturbative QCD at the chiral symmetry breaking scale $\Lambda_\chi\simeq 4\pi F_\pi$ that exhibit themselves as
the poorly-constrained low-energy constants (LECs) in the theory~\cite{Urech:1994hd,Knecht:1999ag}. These natural
limitations prohibit further improvements of the precision level within the original framework.

In this letter we report a new calculation of the EWRC in $K_{e3}$.
Based on a newly-proposed computational framework~\cite{Seng:2019lxf,Seng:2020jtz} that hybridizes the classical
approach by Sirlin~\cite{Sirlin:1977sv} and modern ChPT, we effectively resum the numerically largest terms in the
EWRC to all orders in the chiral expansion and significantly reduce the $\mathcal{O}(e^2p^4)$ uncertainty. Also, we utilize the high-precision lattice QCD calculations of the forward axial $\gamma W$-box
diagrams~\cite{Feng:2020zdc,Ma:2021azh} to constrain the physics from the non-perturbative QCD. With these improvements, we
reduce the theory uncertainty in the EWRC to $K_{e3}$ to an unprecedented level of $10^{-4}$. We will outline here
the most important steps that lead to the final results, while the full detail of the calculation will appear in a
longer paper~\cite{Seng:2021wcf}.

Our primary goal is to study the fractional correction to the $K_{e3}$ decay rate due to EWRC:
\begin{equation}
\delta_{K_{e3}}=\delta \Gamma_{K_{e3}}/\left(\Gamma_{K_{e3}}\right)_\mathrm{tree}\label{eq:deltaKe3}
\end{equation}
up to the precision level of $10^{-4}$. The denominator in Eq.\eqref{eq:deltaKe3} comes from the 
tree-level amplitude for $K(p)\rightarrow\pi(p')e^+(p_e)\nu(p_\nu)$:
\begin{equation}
M_0=-\sqrt{2}G_F\bar{u}_{\nu L}\gamma_\lambda v_{eL}F^\lambda(p',p)~,
\end{equation}
where $G_F$ is the Fermi constant,
$F^\lambda(p',p)=V_{us}^*\left[f_+(t)(p+p')^\lambda+ f_-(t)(p-p')^\lambda\right]$ is the charged weak matrix
element and $f_\pm(t)$ are the charged weak form factors, with $t=(p-p')^2$. We restrict ourselves to $K_{e3}$
for which the contribution from $f_-$ to the decay rate is suppressed by $m_e^2/M_K^2\approx 10^{-6}$ and can be neglected. 

\begin{figure}
	\begin{centering}
		\includegraphics[scale=0.13]{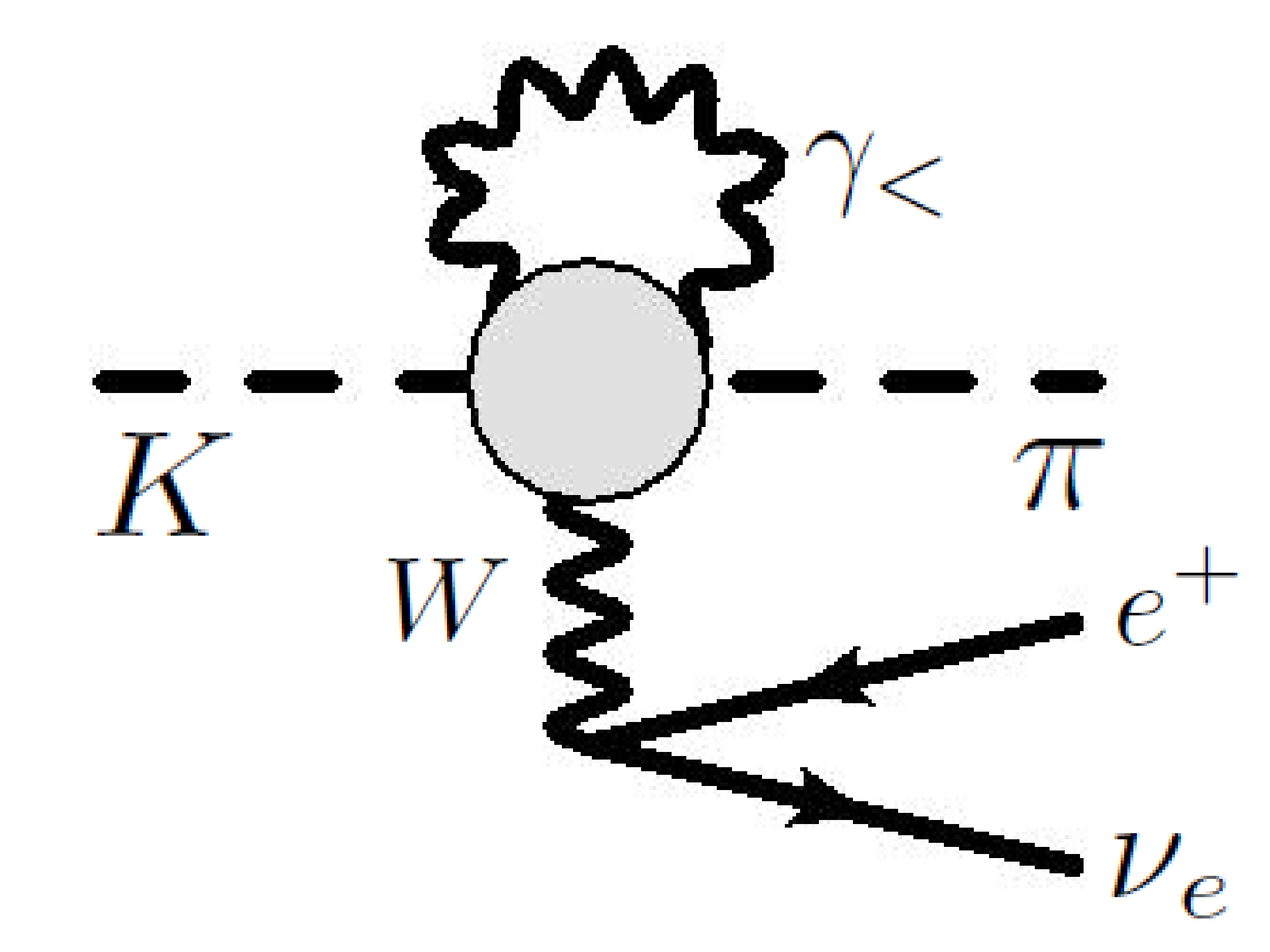}
		\includegraphics[scale=0.13]{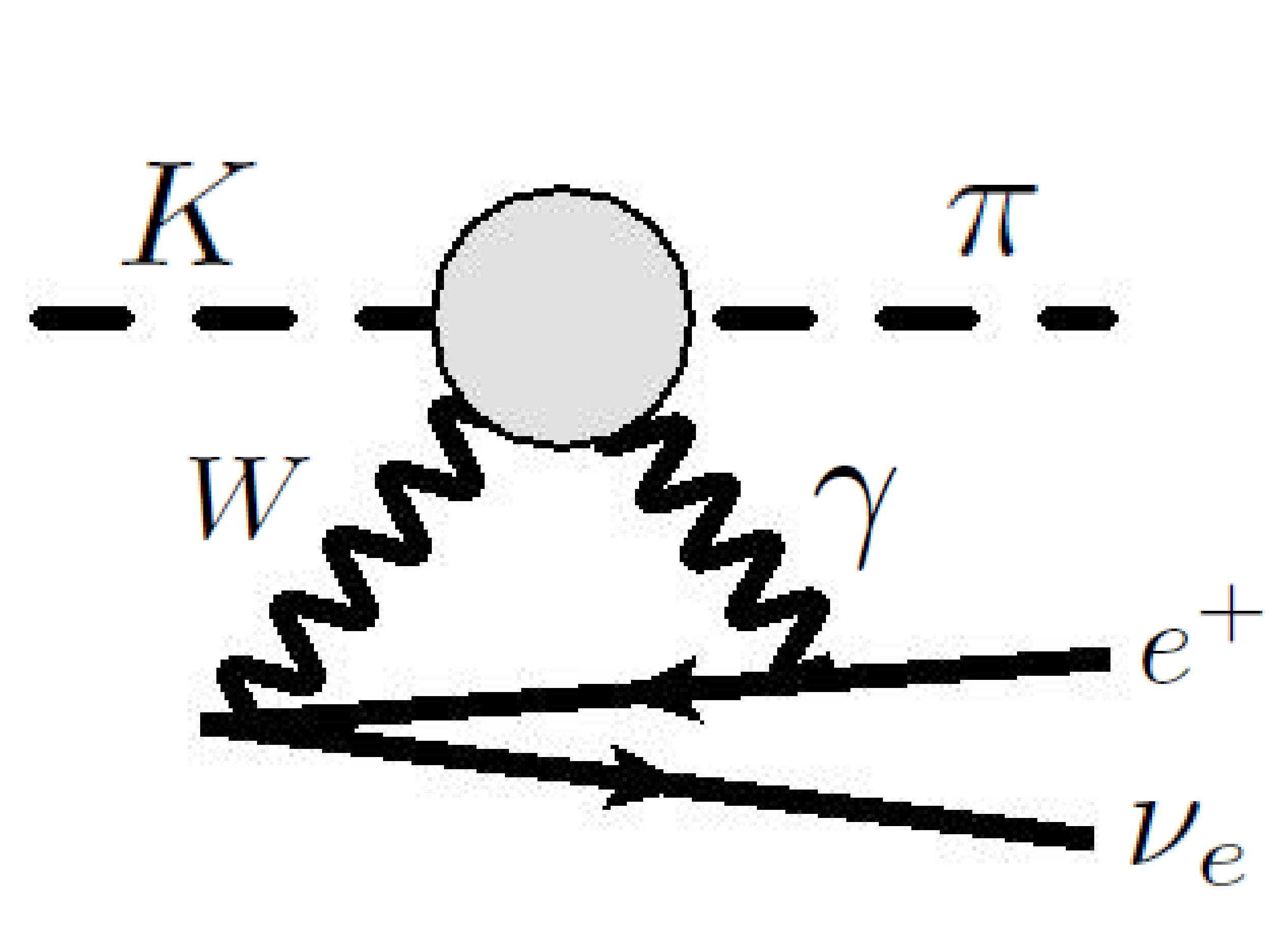}\hfill
		\par\end{centering}
	        \caption{\label{fig:loops}Non-trivial loop diagrams in the $K_{e3}$ EWRC. A factor $M_W^2/(M_W^2-q^{\prime 2})$
                  is attached to the propagator of $\gamma_<$.}
\end{figure}

The full EWRC includes both the virtual corrections and the bremsstrahlung contributions, and we shall start with
the former. A generic one-loop correction to the decay amplitude reads:
\begin{equation}
\delta M_\mathrm{vir}=-\sqrt{2}G_F\bar{u}_{\nu L}\gamma_\lambda v_{eL}I^\lambda~,
\end{equation}
where the loop integrals are contained in $I^\lambda$. It results in a shift of the form factors: $f_\pm\to f_\pm+
\delta f_\pm$, except that $\delta f_\pm$ can also depend on $s=(p'+p_e)^2$ or $u=(p-p_e)^2$. Again, in $K_{e3}$
only $\delta f_+$ is relevant. 

We follow the categorization of the different components of the $\mathcal{O}(G_F\alpha)$ virtual corrections
in Refs.\cite{Seng:2019lxf,Seng:2020jtz}, with $\alpha=e^2/4\pi$.
First, there are pieces in which the loop integrals are independent of
the hadron properties and can be computed analytically. They are contained in Eqs.(2.4) and (2.13) in
Ref.~\cite{Seng:2020jtz}, which combine to give: 
\begin{eqnarray}
\left(\delta f_+\right)_\mathrm{I}&&=\biggl\{\frac{\alpha}{2\pi}\biggl[\ln\frac{M_Z^2}{m_e^2}-\frac{1}{4}\ln\frac{M_W^2}{m_e^2}+\frac{1}{2}\ln\frac{m_e^2}{M_\gamma^2}-\frac{3}{8}\biggr.\biggr.\nonumber\\
\biggl.\biggl.&&\qquad\qquad +\frac{1}{2}\tilde{a}_g\biggr]+\frac{1}{2}\delta_\mathrm{HO}^\mathrm{QED}\biggr\}f_+(t)~,
\end{eqnarray}
where $\tilde{a}_g=-0.083$ and $\delta_\mathrm{HO}^\mathrm{QED}=0.0010(3)$ come from perturbative QCD corrections and
the resummation of large QED logarithms, respectively. Notice also that we have introduced a small photon mass
$M_\gamma$ to regularize the infrared (IR)-divergence.

The remaining loop diagrams in the EWRC, in which the entire dependence
on hadronic structure is contained, are depicted in Fig.\ref{fig:loops}. They depend on the following quantities:
\begin{eqnarray}
T^{\mu\nu}\equiv\int d^4xe^{iq'\cdot x}\left\langle\pi(p')\right|T\{J_\mathrm{em}^\mu(x)J_W^{\nu\dagger}(0)\}\left|K(p)\right\rangle&&\nonumber\\
\Gamma^\mu\equiv \int d^4xe^{iq'\cdot x}\left\langle\pi(p')\right|T\{J_\mathrm{em}^\mu(x)\partial\cdot J_W^{\dagger}(0)\}\left|K(p)\right\rangle~,&&
\end{eqnarray}
which are both functions of the momenta $\{q',p',p\}$.
In particular, we may split the tensor $T^{\mu\nu}$ into two pieces:
$T^{\mu\nu}=\left(T^{\mu\nu}\right)_V+\left(T^{\mu\nu}\right)_A$ that contain the vector and axial component of
the charged weak current, respectively. With these,  the first relevant integral can be written as:
\begin{eqnarray}
&&I_\mathfrak{A}^\lambda =-e^2\int\frac{d^4q'}{(2\pi)^4}\frac{1}{\left[(p_e-q')^2-m_e^2\right]\left[q^{\prime 2}-M_\gamma^2\right]}\nonumber\\
&&\times\biggl\{\frac{2p_e\cdot q'q^{\prime\lambda}}{q^{\prime 2}-M_\gamma^2}T^\mu_{\:\mu}+2p_{e\mu}T^{\mu\lambda}-(p-p')_\mu T^{\lambda\mu}+i\Gamma^\lambda\biggr.\nonumber\\
&&\biggl.-i\epsilon^{\mu\nu\alpha\lambda}q_\alpha'\left(T_{\mu\nu}\right)_V\biggr\}~,\label{eq:IA}
\end{eqnarray}
where the first two lines come from Eq.(2.13) of Ref.~\cite{Seng:2020jtz}, and the third line is a part of $\delta M_{\gamma W}^A$ in Eq.~(2.10)
of the same paper.

\begin{figure}
	\begin{centering}
		\includegraphics[scale=0.19]{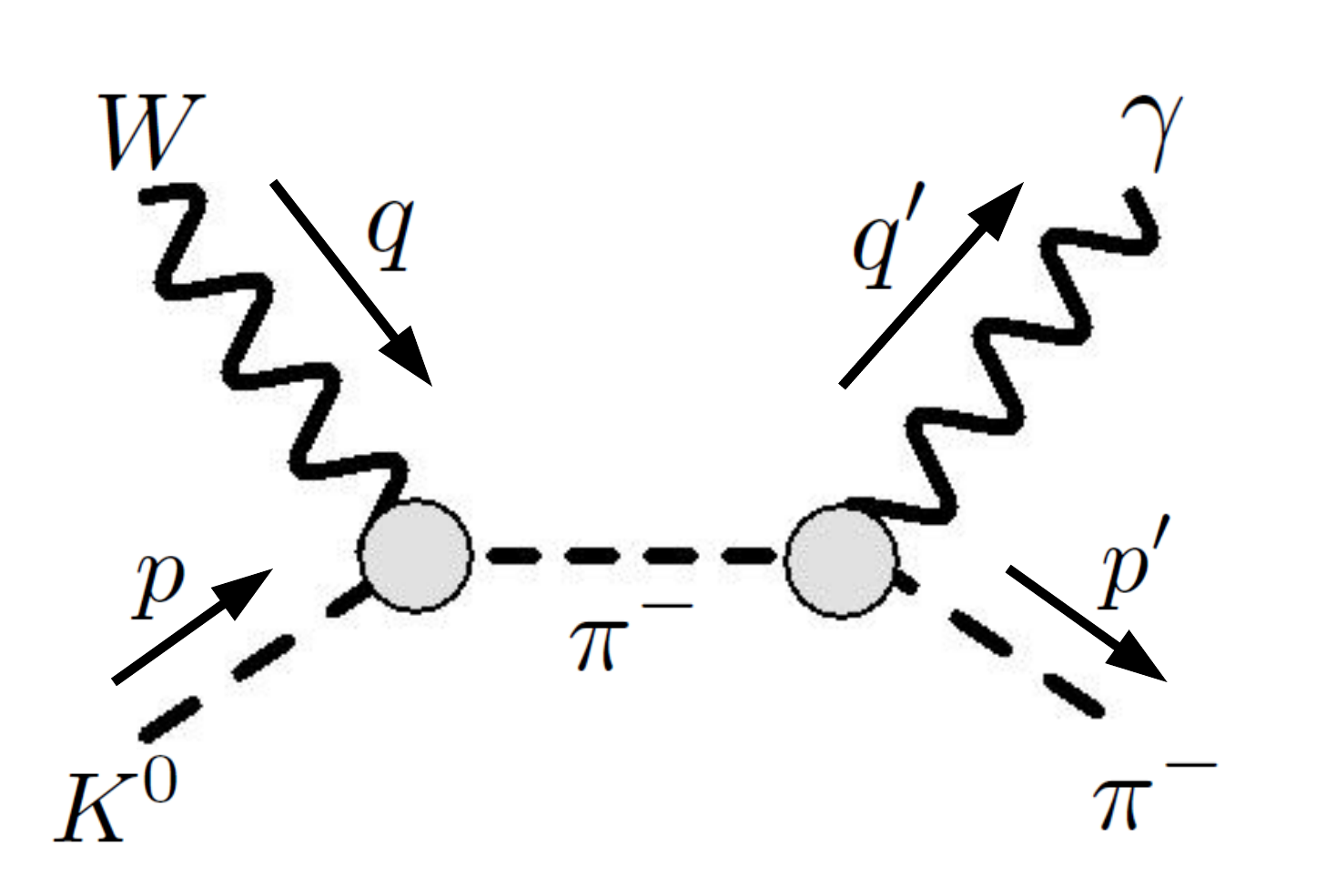}
		\includegraphics[scale=0.19]{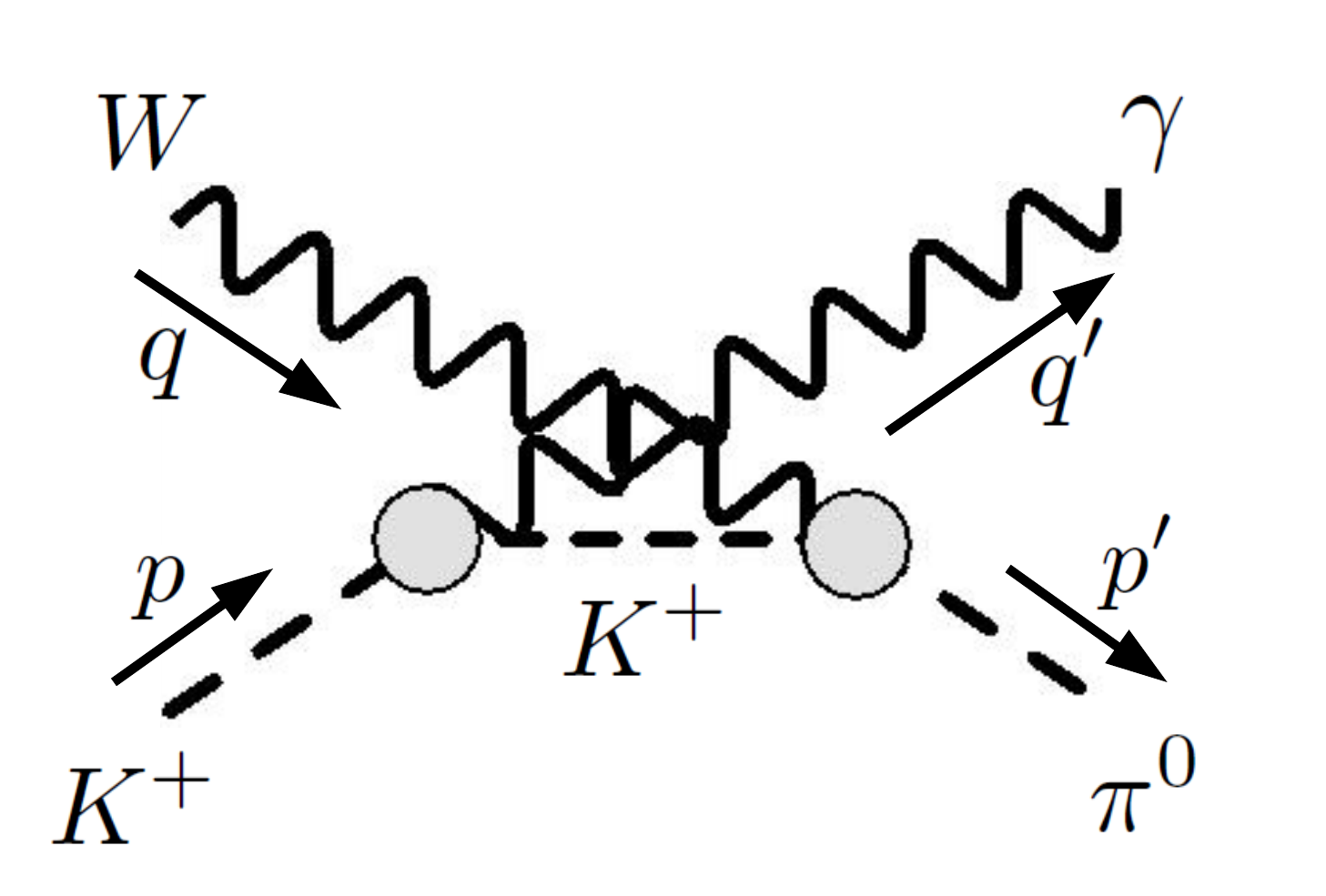}
		\includegraphics[scale=0.19]{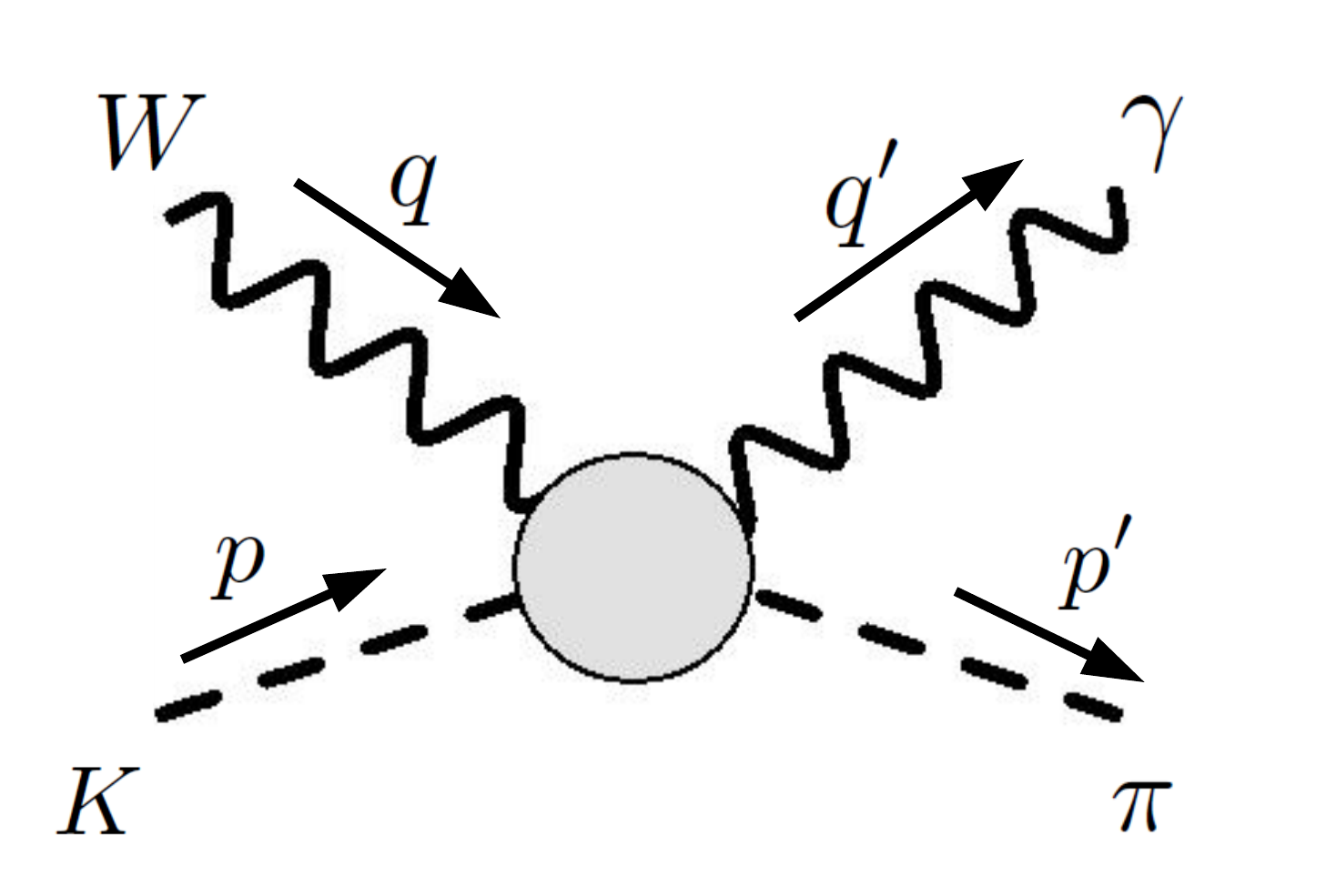}\hfill
		\par\end{centering}
	\caption{\label{fig:TmunulowQ}Pole (left, middle) and seagull diagrams.}
\end{figure}

The operator product expansion (OPE) shows that the $|q'|>\Lambda_\chi$ region does not contribute to the
integral $I_\mathfrak{A}^\lambda$, therefore only the low-energy expressions of $T^{\mu\nu}$ and $\Gamma^\mu$ are needed.
To this end, we find it useful to split them into the ``pole'' and ``seagull'' terms respectively, as
depicted in Fig.~\ref{fig:TmunulowQ}:
\begin{equation}
T^{\mu\nu}=T^{\mu\nu}_\mathrm{pole}+T^{\mu\nu}_\mathrm{sg},\:\:\Gamma^\mu=\Gamma^\mu_\mathrm{pole}+\Gamma^\mu_\mathrm{sg}~.
\end{equation}
Furthermore, we can obtain the so-called ``convection term'' by setting $q'\to 0$ in both the electromagnetic
form factor and the charged weak vertex of the pole term~\cite{Meister:1963zz}. It represents the minimal
expression that satisfies the exact electromagnetic Ward identity, and thus gives the full IR-divergent structures
in the loop integrals.

The seagull term receives contributions from resonances and the many-particle continuum. An 
estimate operating with low-lying resonances~\cite{Ecker:1988te,Ecker:1989yg,Cirigliano:2006hb} suggests that its
contribution to $\delta_{K_{e3}}$ is at most $10^{-4}$. Note that t-channel exchanges that still retain some sensitivity to the long-range effects do not contribute to Eq.\eqref{eq:IA}. To stay on the conservative side, we assign to 
it a generic uncertainty of $2\times 10^{-4}$. Therefore, $\delta f_+$ derived from $I_\mathfrak{A}^\lambda$ is dominated by the pole contribution
which is fully determined by the $K$ and $\pi$ electromagnetic and charged weak form factors. The result splits
into two pieces:  
\begin{equation}
\left(\delta f_+\right)_\mathfrak{A}=\left(\delta f_+\right)_\mathrm{II}+\left(\delta f_+\right)_\mathfrak{A}^\mathrm{fin}~,\label{eq:deltafA}
\end{equation}
where $\left(\delta f_+\right)_\mathrm{II}$ is a model-independent IR-divergent piece. The IR-finite piece,
$\left(\delta f_+\right)_\mathfrak{A}^\mathrm{fin}$, on the other hand, is evaluated numerically by adopting a monopole
parameterization of the hadronic form factors~\cite{Amendolia:1986ui,Amendolia:1986wj,Lazzeroni:2018glh}. Notice
that the integral $I_\mathfrak{A}^\lambda$ only probes the region $q'\sim p_e\sim p-p'\sim M_K-M_\pi$, where different
parameterizations of the form factors are practically indistinguishable. In particular, we find that the main source of the
uncertainty is the $K^+$ mean-square charge radius and the experimental uncertainty thereof,
$\left\langle r_K^2\right\rangle=0.34(5)$~fm$^2$~\cite{Amendolia:1986ui}.

The second relevant integral is:
\begin{equation}
  I_\mathfrak{B}^\lambda=ie^2\int\frac{d^4q'}{(2\pi)^4}\frac{M_W^2}{M_W^2-q^{\prime 2}}
  \frac{\epsilon^{\mu\nu\alpha\lambda}q_\alpha'\left(T_{\mu\nu}\right)_A}{\left[(p_e-q')^2-m_e^2\right]q^{\prime 2}}~,\label{eq:IB}
\end{equation}
which picks up the remaining part of $\delta M_{\gamma W}^A$ in Eq.(2.10) of Ref.~\cite{Seng:2020jtz}. It is IR-finite,
but probes the physics from $|q'|=0$ all the way up to $|q'|\sim M_W$. A significant amount of theoretical uncertainty
thus resides in the region $|q'|\sim\Lambda_\chi$ where non-perturbative QCD takes place, and has been an unsettled
issue for decades. The situation is changed following the recent lattice QCD calculations of the so-called ``forward
axial $\gamma W$-box'':
\begin{eqnarray}
&&\Box_{\gamma W}^{VA}(\phi_i,\phi_f,M)\equiv\frac{ie^2}{2M^2}\int\frac{d^4q'}{(2\pi)^4}\frac{M_W^2}{M_W^2-q^{\prime 2}}\frac{\epsilon^{\mu\nu\alpha\beta}q_\alpha'p_\beta}{(q^{\prime 2})^2}\nonumber\\
&&\times\frac{T_{\mu\nu}^{if}(q',p,p)}{F_+^{if}(0)}~,\label{eq:axialgammaW}
\end{eqnarray}
where $T_{\mu\nu}^{if}$ is just $T_{\mu\nu}$ except that the initial and final states are now $\{\phi_i,\phi_f\}$ with $p^2=M_i^2=M_f^2=M^2$, and $F_+^{if}(0)$ is the form factor $f_+^{if}(0)$ multiplied by the appropriate CKM matrix element. Following the existing literature, we split it into two pieces:
\begin{equation}
\Box_{\gamma W}^{VA}(\phi_i,\phi_f,M)=\Box_{\gamma W}^{VA>}+\Box_{\gamma W}^{VA<}(\phi_i,\phi_f,M)
\end{equation}
which come from the loop integral at $Q^2\equiv -q^{\prime 2}>Q_\mathrm{cut}^2$ and $Q^2<Q_\mathrm{cut}^2$ respectively, where $Q_\mathrm{cut}^2=2$~GeV$^2$ is a scale above which perturbative QCD works well. The ``$>$'' term is flavor- and mass-independent, and was calculated to $\mathcal{O}(\alpha_s^4)$: $\Box_{\gamma W}^{VA>}=2.16\times 10^{-3}$~\cite{Feng:2020zdc}. In the meantime, direct lattice calculations of the ``$<$'' term were performed in two channels~\cite{Feng:2020zdc,Ma:2021azh}:
\begin{eqnarray}
\Box_{\gamma W}^{VA<}(\pi^+,\pi^0,M_\pi)&=&0.67(3)_\mathrm{lat}\times 10^{-3}\nonumber\\
\Box_{\gamma W}^{VA<}(K^0,\pi^-,M_\pi)&=&0.28(4)_\mathrm{lat}\times 10^{-3}
\end{eqnarray}
from which we can also obtain $\Box_{\gamma W}^{VA<}(K^+,\pi^0,M_\pi)=1.06(7)_\mathrm{lat}\times 10^{-3}$ through a ChPT matching~\cite{Seng:2020jtz}.

The only difference between the integrals in Eq.\eqref{eq:IB} and \eqref{eq:axialgammaW} is the non-forward (NF) kinematics in the former (i.e. $p\neq p'$ and $p_e\neq 0$), which only affect the integral in the $Q^2<Q_\mathrm{cut}^2$ region. Therefore one could similarly split $\left(\delta f_+\right)_\mathfrak{B}$ into two pieces: $\left(\delta f_+\right)_\mathfrak{B}=\left(\delta f_+\right)_\mathfrak{B}^>+\left(\delta f_+\right)_\mathfrak{B}^<$, where the ``$>$'' piece matches trivially to the forward axial $\gamma W$-box:
\begin{equation}
\left(\delta f_+\right)_\mathfrak{B}^>=\Box_{\gamma W}^{VA>}f_+(t)~.
\end{equation}
On the other hand, the matching between the ``$<$'' components is not exact due to the NF effects. We characterize the latter by an energy scale $E$ that could be either $M_K-M_\pi$, $(s-M_\pi^2)^{1/2}$ or $(u-M_\pi^2)^{1/2}$. The matching then reads:
\begin{equation}
\left(\delta f_+\right)_\mathfrak{B}^<=\left\{\Box_{\gamma W}^{VA<}(K,\pi,M_\pi)+\mathcal{O}\left(E^2/\Lambda_\chi^2\right)\right\}f_+(t)~,\label{eq:Boxsmallmatch}
\end{equation}
where $\mathcal{O}(E^2/\Lambda_\chi^2)$ represents the NF corrections. Numerically, since $E<M_K$, we may multiply the right-hand side of Eq.~\eqref{eq:Boxsmallmatch} by $M_K^2/\Lambda_\chi^2$ as a conservative estimation of the NF uncertainty. 

The last virtual correction is the so-called ``three-point function'' contribution to the charged weak form factors, which was derived within ChPT to $\mathcal{O}(e^2p^2)$ in Ref.~\cite{Seng:2019lxf}. However, it contains an IR-divergent piece that comes from the convection term contribution, and can be resummed to all orders in the chiral expansion by simply adding back the charged weak form factors. This leads to the following partially-resummed ChPT expression:
\begin{equation}
\delta f_{+,3}=\left(\delta f_+\right)_\mathrm{III}+\left\{\left(\delta f_{+,3}\right)_{e^2p^2}^\mathrm{fin}+\mathcal{O}(e^2p^4)\right\}~,
\end{equation}
where the IR-divergent piece $\left(\delta f_+\right)_\mathrm{III}$ is exact, i.e. resummed to all orders in ChPT. It combines with $\left(\delta f_+\right)_\mathrm{II}$ in Eq.~\eqref{eq:deltafA} to give:
\begin{eqnarray}
\mathfrak{Re}\left(\delta f_+\right)_\mathrm{II+III}&=&\frac{\alpha}{4\pi}\left[-\frac{2}{\beta_i}\tanh^{-1}\beta_i\ln\left(\frac{M_im_e}{M_\gamma^2}\right)\right.\nonumber\\
&&\left.+\ln\frac{M_i^2}{M_\gamma^2}-\frac{5}{2}\right]f_+(t)~,
\end{eqnarray}
where $M_i$ is the mass of the charged meson ($K^+$ in $K_{e3}^+$ and $\pi^-$ in $K_{e3}^0$) and $\beta_i$ is the speed of the positron in the rest frame of the charged meson. Meanwhile, the IR-finite pieces, $\left(\delta f_{+,3}\right)_{e^2p^2}^\mathrm{fin}$, are given by the terms in Eqs.~(8.3) and (8.5) of Ref.~\cite{Seng:2019lxf} that are not attached to the factor $\ln(M^2/M_\gamma^2)-5/2$, and are subject to $\mathcal{O}(e^2p^4)$ corrections. 

\begin{figure}
	\begin{centering}
		\includegraphics[scale=0.25]{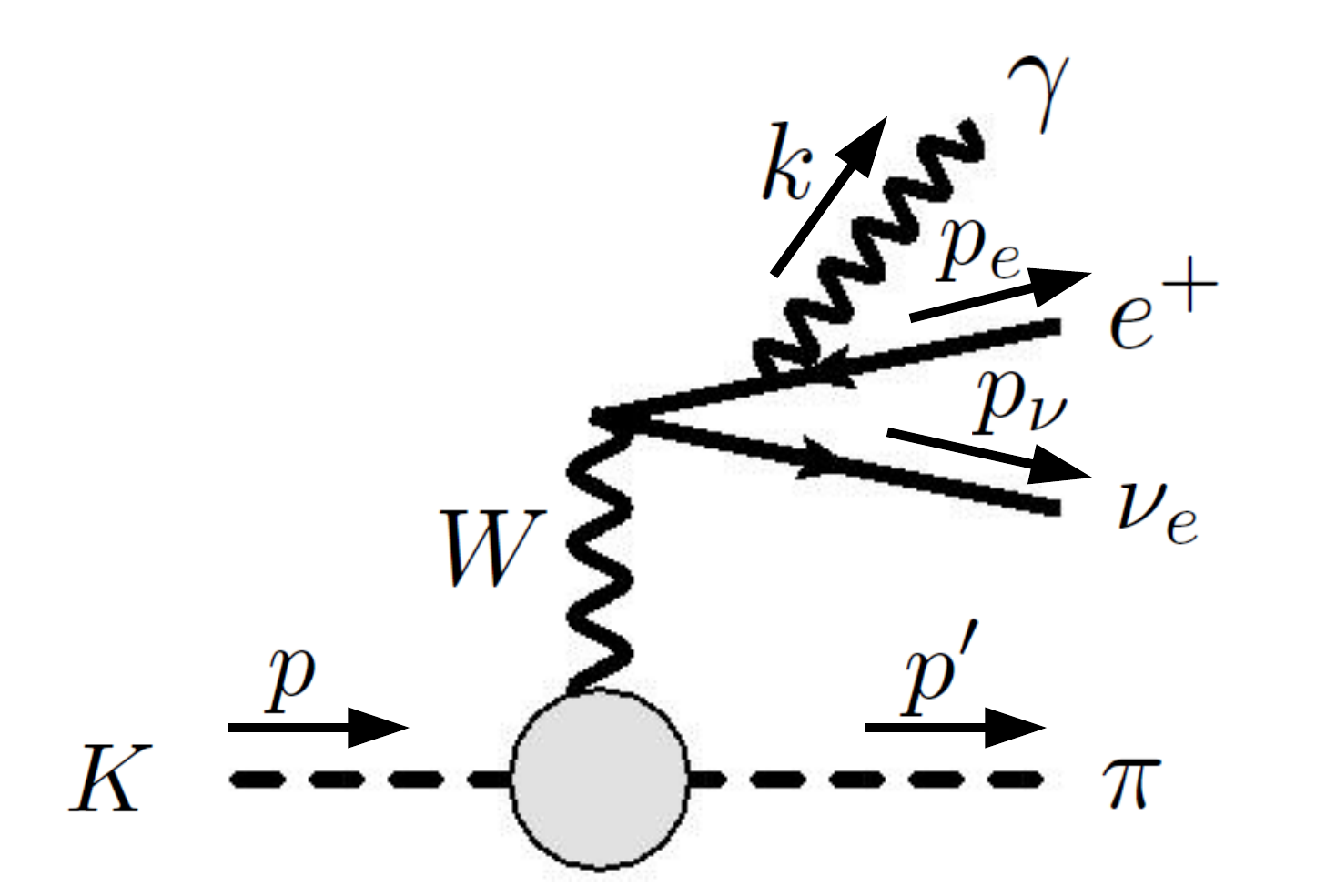}	\includegraphics[scale=0.25]{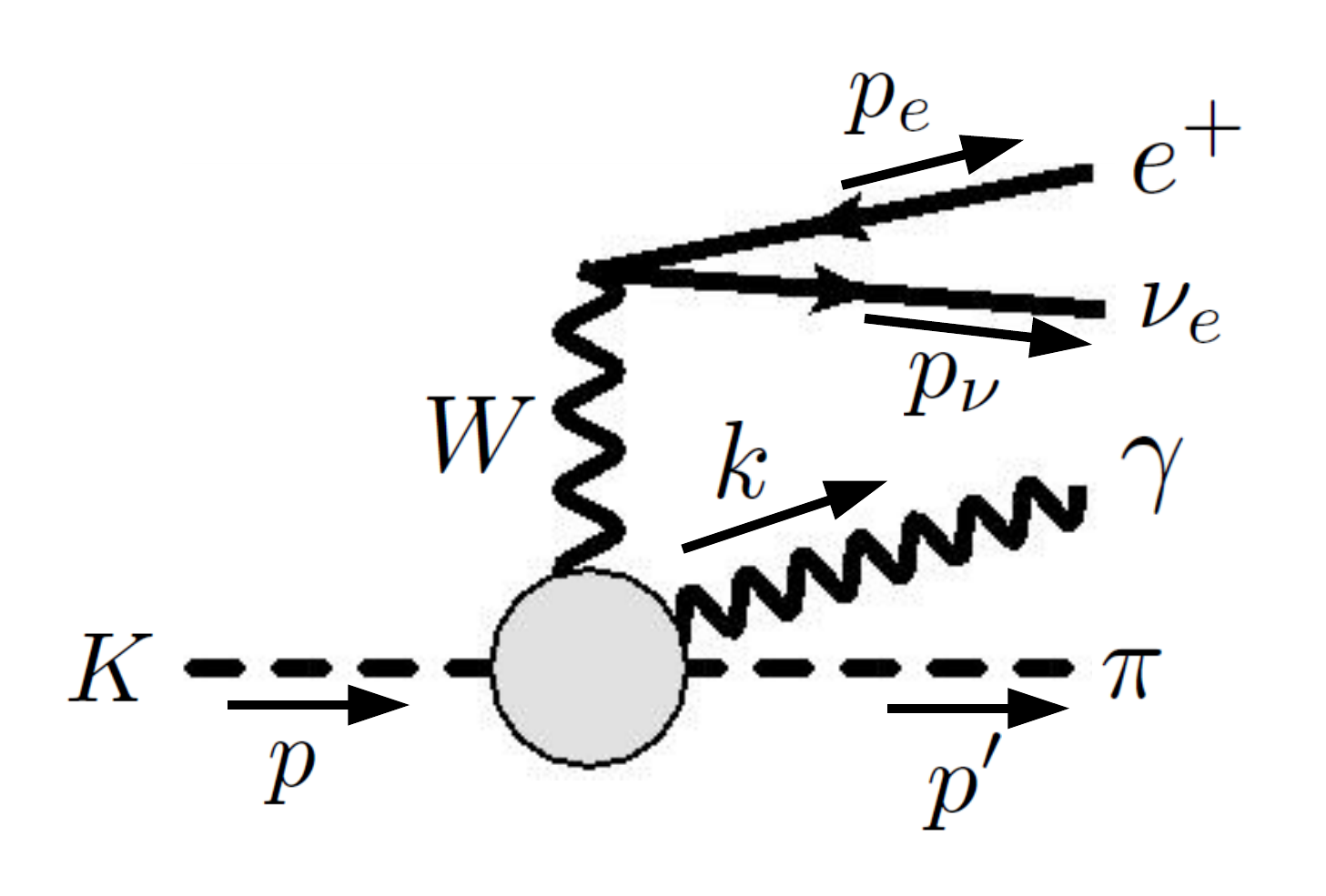}\hfill
		\par\end{centering}
	\caption{\label{fig:brems}Bremsstrahlung diagrams.}
\end{figure}

\begin{table*}[t]
	\begin{centering}
		\begin{tabular}{|c|c|c|c|c|}
			\hline 
			& from $\left(\delta f_{+}\right)_{\mathfrak{A}}^{\mathrm{fin}}$ & from $\left(\delta f_{+}\right)_{\mathfrak{B}}$ & from $\left(\delta f_{+}\right)_{\mathrm{I+II+III}}$ and $|M_{A}|^{2}$ & from $2\mathfrak{Re}\left\{ M_{A}^{*}M_{B}\right\} +|M_{B}|^{2}$\tabularnewline
			\hline 
			\hline 
			$K_{e3}^{0}$ & $-0.09(2)_{\mathrm{sg}}$ & $0.49(1)_{\mathrm{lat}}(1)_{\mathrm{NF}}$ & $2.97(3)_{\mathrm{HO}}$ & $0.12(2)_{e^{2}p^{4}}$\tabularnewline
			\hline 
			$K_{e3}^{+}$ & $0.96(2)_{\mathrm{sg}}(1)_{\left\langle r_{K}^{2}\right\rangle }$ & $0.64(1)_{\mathrm{lat}}(4)_{\mathrm{NF}}$ & $0.97(3)_{\mathrm{HO}}$ & $-0.04(1)_{e^{2}p^{4}}$\tabularnewline
			\hline 
		\end{tabular}
		\par\end{centering}
	\caption{Summary of various contributions to $\delta_{K_{e3}}$ (except that from $\left(\delta f_{+,3}\right)_{e^2p^2}^\mathrm{fin}$, see the discussions after Eq.\eqref{eq:GammaKe3}) in units of
		$10^{-2}$.\label{tab:summary}}	
\end{table*}

Next we switch to the bremsstrahlung contributions, as depicted
in Fig.~\ref{fig:brems}. Its amplitude is given by:
\begin{eqnarray}
&&M_\mathrm{brems}=-\sqrt{2}G_Fe\bar{u}_{\nu L}\gamma^\mu\left\{\frac{p_e\cdot \varepsilon^*}{p_e\cdot k}+\frac{\slashed{k}\slashed{\varepsilon}^*}{2p_e\cdot k}\right\}v_{eL}F_\mu\nonumber\\
&&+i\sqrt{2}G_Fe\bar{u}_{\nu L}\gamma^\nu v_{eL}\varepsilon^{\mu*}T_{\mu\nu}(k;p',p)~,
\end{eqnarray}
in which the tensor $T^{\mu\nu}$ appears again, except that now it deals with an on-shell photon momentum $k$ whose size is restricted by phase space. Similar to $\delta f_{+,3}$, we find that the most efficient way to calculate the bremsstrahlung contributions is to adopt a partially-resummed ChPT expression for $T^{\mu\nu}$:
\begin{equation}
T^{\mu\nu}=T^{\mu\nu}_\mathrm{conv}+\left\{\left(T^{\mu\nu}-T^{\mu\nu}_\mathrm{conv}\right)_{p^2}+\mathcal{O}(p^4)\right\}~,\label{eq:Tmunubrem}
\end{equation}
where the full convection term $T_\mathrm{conv}^{\mu\nu}$ is explicitly singled out, while the remaining terms in the curly bracket are expanded to $\mathcal{O}(p^2)$. Consequently, one can split $M_\mathrm{brems}$
into two separately gauge-invariant pieces:
\begin{equation}
M_\mathrm{brems}=M_A+M_B~,
\end{equation} 
where the terms in the curly bracket of Eq.\eqref{eq:Tmunubrem} reside in $M_B$. The contribution to the decay rate from $|M_A|^2$ contains the full IR-divergent structure (which cancels with the virtual corrections), is numerically the largest and does not associate to any chiral expansion uncertainty. The term $2\mathfrak{Re}\left\{M_A^*M_B\right\}+|M_B|^2$, on the other hand, is subject to $\mathcal{O}(e^2p^4)$ corrections. We find that its contribution to $\delta_{K_{e3}}$ is $\lesssim 10^{-3}$, so the associated chiral expansion uncertainty, which is obtained by multiplying the central value with $M_K^2/\Lambda_\chi^2$, is of the order $10^{-4}$. 

With the above, we have calculated all EWRC to $10^{-4}$ and may compare with existing results. The standard parameterization of the fully-inclusive $K_{e3}$ decay rate reads~\cite{Zyla:2020zbs}:
\begin{eqnarray}
\Gamma_{K_{e3}}&=&\frac{G_F^2|V_{us}|^2M_K^5C_K^2}{192\pi^3}S_\mathrm{EW}|f_+^{K^0\pi^-}(0)|^2I_{Ke}^{(0)}(\lambda_i)\nonumber\\
&&\times(1+\delta_\mathrm{EM}^{Ke}+\delta_\mathrm{SU(2)}^{K\pi})~,\label{eq:GammaKe3}
\end{eqnarray}
among which $S_\mathrm{EW}=1.0232(3)$ describes the short-distance EWRC~\cite{Marciano:1993sh} (the uncertainty comes from $\delta_\mathrm{HO}^\mathrm{QED}$~\cite{Erler:2002mv}) and $\delta_\mathrm{EM}^{Ke}$ describes the long-distance electromagnetic corrections respectively. We also realize that in the existing ChPT treatment a residual component of the electromagnetic corrections, which corresponds exactly to $\left(\delta f_{+,3}\right)_{e^2p^2}^\mathrm{fin}$ in our language, is redistributed into $I_{Ke}^{(0)}(\lambda_i)$ and $\delta_\mathrm{SU(2)}^{K\pi}$ that describe the $t$-dependence of the charged weak form factors and the isospin breaking correction, respectively~\cite{Cirigliano:2001mk,Cirigliano:2004pv,Cirigliano:2008wn}. Therefore, the correspondence between $\delta_\mathrm{EM}^{Ke}$ in the ChPT calculation and $\delta_{K_{e3}}$ in our approach reads:
\begin{equation}
\delta_\mathrm{EM}^{Ke}=\left(\delta_{K_{e3}}\right)_\mathrm{tot}-\left(S_\mathrm{EW}-1\right)-\left(\delta_{K_{e3}}\right)_3^\mathrm{fin}~.
\end{equation}

Our results of the different components of $\delta_{K_{e3}}$ are summarized in Table~\ref{tab:summary}, from which 
we obtain:
\begin{eqnarray}
\delta_\mathrm{EM}^{K^+e}&=&0.21(2)_\mathrm{sg}(1)_{\left\langle r_K^2\right\rangle}(1)_\mathrm{lat}(4)_\mathrm{NF}(1)_{e^2p^4}\times 10^{-2}\nonumber\\
\delta_\mathrm{EM}^{K^0e}&=&1.16(2)_\mathrm{sg}(1)_\mathrm{lat}(1)_\mathrm{NF}(2)_{e^2p^4}\times 10^{-2}~.\label{eq:deltaEM}
\end{eqnarray}
The uncertainties are explained as follows: ``sg'' is our estimate of the seagull contribution to $I_\mathfrak{A}^\lambda$, ``$\left\langle r_K^2\right\rangle$'' comes from the experimental uncertainty of the $K^+$ mean-square charge radius that enters $I_\mathfrak{A}^\lambda$, ``lat'' and ``NF'' are the uncertainties in $\left(\delta f_+\right)_\mathfrak{B}$ from lattice QCD and the NF effects, respectively, and ``$e^2p^4$'' represents the chiral expansion uncertainty in the $2\mathfrak{Re}\left\{M_A^*M_B\right\}+|M_B^2|$ term from the bremsstrahlung contribution. 
We should compare Eq.\eqref{eq:deltaEM} to the ChPT result~\cite{Cirigliano:2008wn}:
\begin{eqnarray}
\left(\delta_\mathrm{EM}^{K^+e}\right)_\mathrm{ChPT}&=&0.10(19)_{e^2p^4}(16)_\mathrm{LEC}\times 10^{-2}\nonumber\\ \left(\delta_\mathrm{EM}^{K^0e}\right)_\mathrm{ChPT}&=&0.99(19)_{e^2p^4}(11)_\mathrm{LEC}\times 10^{-2}~.
\end{eqnarray}
They are consistent within error bars, but Eq.~\eqref{eq:deltaEM} shows a reduction of the total uncertainty by almost an order of magnitude, which can be easily understood as follows. First, in ChPT the $\mathcal{O}(e^2p^4)$ uncertainty is obtained by multiplying the full result, including the IR-singular pieces that are numerically the largest, with $M_K^2/\Lambda_\chi^2$; meanwhile, within the new formalism those pieces can be evaluated exactly by simply isolating the pole/convection term in $T^{\mu\nu}$ and $\Gamma^\mu$. The remainders are generically an order of magnitude smaller, so their associated $\mathcal{O}(e^2p^4)$ uncertainty is also suppressed. Secondly, in ChPT the LECs $\{X_i\}$ were estimated within resonance models~\cite{Ananthanarayan:2004qk,DescotesGenon:2005pw} and were assigned a 100\% uncertainty. On the other hand, some of us pointed out in Ref.~\cite{Seng:2020jtz} that these LECs are associated with the forward axial $\gamma W$-box diagram, and promoted first-principle calculations with lattice QCD. This effectively transforms the LEC uncertainties in ChPT into the lattice and NF uncertainties in $\left(\delta f_+\right)_\mathfrak{B}$ which are much better under control.  

To conclude, we performed a significantly improved calculation of the EWRC in the $K_{e3}$ channel. We observe no large systematic corrections with respect to previous analyses. Although the error analysis in the $K_{\mu 3}$ channel is somewhat more complicated, we deem such large corrections in this channel unlikely. Hence, it is safe to conclude that the EWRC in $K_{l3}$ cannot be responsible for the $K_{\mu 2}$--$K_{l3}$ discrepancy in $V_{us}$. One should then switch to other SM inputs, such as the lattice calculation of $|f_+^{K^0\pi^-}(0)|$ and the theory inputs of $I_{Kl}^{(0)}(\lambda_i)$ and $\delta_\mathrm{SU(2)}^{K\pi}$. 
Finally, our improvement in $\delta_\mathrm{EM}^{Ke}$ also opens a new pathway for the precise measurement of $V_{us}/V_{ud}$ through the ratio between the semileptonic kaon and pion decay rate~\cite{Czarnecki:2019iwz}. For instance, we may define:
\begin{equation}
R_V\equiv\frac{\Gamma_{K_{e3}^0}}{\Gamma_{\pi_{e3}}}=4.035(4)_\mathrm{PS}(1)_\mathrm{RC}\times 10^8\left|\frac{V_{us}f_+^{K^0\pi^-}(0)}{V_{ud}f_+^{\pi^+\pi^0}(0)}\right|^2.
\end{equation}	
Since both the RC uncertainties in $K_{e3}^0$ and $\pi_{e3}$ are now at the $10^{-4}$ level, the dominant theory uncertainty (apart from lattice inputs) of $R_V$ comes from the $K_{e3}^0$ phase space (PS) integral. We compare this to:
\begin{equation}
R_A\equiv \frac{\Gamma_{K_{\mu 2}}}{\Gamma_{\pi_{\mu 2}}}=17.55(3)_\mathrm{RC}\left|\frac{V_{us}f_{K^+}}{V_{ud}f_{\pi^+}}\right|^2
\end{equation}
which is currently used to extract $V_{us}/V_{ud}$. We see that $R_V$ possesses a much smaller theoretical uncertainty than $R_A$, and hence represents a more promising avenue in the future. Our work thus provides a strong motivation for experimentalists to measure the $\pi_{e3}$ branching ratio with an order-of-magnitude increase in precision~\cite{ArevaloSnowmass}.

\begin{acknowledgments}
We thank Vincenzo Cirigliano for many inspiring discussions. This work is supported in
part by  the DFG (Projektnummer 196253076 - TRR 110)
and the NSFC (Grant No. 11621131001) through the funds provided
to the Sino-German TRR 110 ``Symmetries and the Emergence of
Structure in QCD" (U-G.M, C.Y.S and D.G), by the Alexander von Humboldt Foundation through the Humboldt
Research Fellowship (C.Y.S), by the Chinese Academy of Sciences (CAS) through a President's
International Fellowship Initiative (PIFI) (Grant No. 2018DM0034) and by the VolkswagenStiftung
(Grant No. 93562) (U-G.M), by EU Horizon 2020 research and innovation programme, STRONG-2020 project
under grant agreement No 824093 and by the German-Mexican research collaboration Grant No. 278017 (CONACyT)
and No. SP 778/4-1 (DFG) (M.G). 
\end{acknowledgments}

\end{document}